\documentclass{aa}
\usepackage{natbib}
\usepackage{graphicx}
\usepackage{txfonts}
\bibpunct{(}{)}{;}{a}{}{,}
\usepackage{hyperref}
\usepackage[para,online,flushleft]{threeparttable}
\usepackage{cprotect}
\graphicspath{ {images/} }
\makeatletter
\renewcommand*\aa@pageof{, page \thepage{} of \pageref*{LastPage}}
\makeatother
\begin{document} 
  \title{Planetary mass-radius relations across the galaxy}
  \author{
  A. Michel\inst{1,2}, J. Haldemann\inst{1}, C. Mordasini\inst{1},\and Y. Alibert\inst{1}
  }
% ----------------------------------------------------------------------------------------  
%Author ORCIDs
%Arnaud: https://orcid.org/0000-0003-4099-9026
%Jonas: https://orcid.org/0000-0003-1231-2389
%Christoph:https://orcid.org/0000-0002-1013-2811
%Yann:https://orcid.org/0000-0002-4644-8818
% ----------------------------------------------------------------------------------------  
  \institute{Physikalisches Institut, Universität Bern, Gesellschaftsstrasse 6, 3012 Bern, Switzerland \\  \email{arnaud.michel@questu.ca}
  \and
  Quest University Canada, 3200 University Boulevard, Squamish, V8B 0N8, BC, Canada}

  \date{Received 15 October 2020 / Accepted 25 May 25 2020}
% ----------------------------------------------------------------------------------------
  \abstract
  % context heading 
   {Planet formation theory suggests that planet bulk compositions are likely to reflect the chemical abundance ratios of their host star's photosphere. Variations in the abundance of particular chemical species in stellar photospheres between different galactic stellar populations demonstrate that there are differences among the expected solid planet bulk compositions.}
   {We aim to present planetary mass-radius relations of solid planets for kinematically differentiated stellar populations, namely, the thin disc, thick disc, and halo.}
   {Using two separate internal structure models, we generated synthetic planets using bulk composition inputs derived from stellar abundances. We explored two scenarios, specifically iron-silicate planets at 0.1 AU and silicate-iron-water planets at 4 AU.}
   {We show that there is a persistent statistical difference in the expected mass-radius relations of solid planets among the different galactic stellar populations. At 0.1 AU for silicate-iron planets, there is a 1.51 to 2.04\% mean planetary radius difference between the thick and thin disc stellar populations, whilst for silicate-iron-water planets past the ice line at 4 AU, we calculate a 2.93 to 3.26\% difference depending on the models. Between the halo and thick disc, we retrieve at 0.1 AU a 0.53 to 0.69\% mean planetary radius difference, and at 4 AU we find a 1.24 to 1.49\% difference depending on the model.}
   {Future telescopes (such as PLATO) will be able to precisely characterize solid exoplanets and demonstrate the possible existence of planetary mass-radius relationship variability between galactic stellar populations.}
% ----------------------------------------------------------------------------------------
% ----------------------------------------------------------------------------------------
\authorrunning{A. Michel et al.}
\titlerunning{Planetary \textit{M - R} relations across the galaxy}
\maketitle
% ----------------------------------------------------------------------------------------
\section{Introduction}
% ----------------------------------------------------------------------------------------
   Exoplanet surveys are beginning to yield large amounts of observational data on low-mass exoplanets \citep{Batalha_2013, Mayor_2014, Dressing_Charbonneau_2015}. There is substantial evidence from planet formation theory and condensation models suggesting that particular chemical species present in a host star’s photosphere are reflected in the composition of those planets \citep{Bond_2010, DelgadoMena_2010, Thiabaud_2015, Dorn_2017_1, Dorn_2017_2, Santos_2017}. The correlation exists in the solar system where there is good agreement between the chemical abundance ratios in the solar photosphere and primitive meteorites, the latter is often used to ascribe rocky planet bulk compositions \citep{Lodders_2003}.

   The solar neighbourhood in our galaxy is composed of three distinct stellar populations, the thin disc (of which the Sun is a part), the thick disc, and the halo \citep{Buser_2000, Nissen_2004}. These three galactic substructures can be separated according to their kinematics \citep{Reddy_2006, Bensby_2014}, age \citep{Bensby_2005, Yoachim_2008}, chemical abundances \citep{Navarro_2011, Adibekyan_2012_1}, or a combination of the three \citep{Fuhrmann_1998, Haywood_2013}.

   From \citet{Santos_2017} and \citet{Cabral_2019}, the three stellar populations demonstrate variability in the chemical abundance ratios that are expected to form the bulk composition of rocky planets. In these studies, observations from a HARPS sample from \citet{Adibekyan_2012_1} are used by \citet{Santos_2017}. Separately, the Besançon stellar population synthesis model \citep{Lagarde_2017} is used by \citet{Cabral_2019} to show the expected different iron and water mass fractions between galactic stellar populations.

   Within this context, we investigate whether we can differentiate the mass-radius (\textit{M - R}) relations for solid planets between the three galactic stellar populations using stellar chemical compositions. The paper is organized as follows; in Sect.~\ref{sec:data}, we present the stellar sample and the kinematic separation method. In Sect.~\ref{sec:methods}, we discuss the two internal structure models used to generate synthetic solid planets. In this same section, we also explain the analytic stoichiometric model used to determine the planetary composition inputs to the models. In Sect.~\ref{sec:results}, we present our results. In Sect.~\ref{sec:discussion}, we discuss and place the results in perspective of possible observations.
% ----------------------------------------------------------------------------------------
\section{Data}
\label{sec:data}
% ----------------------------------------------------------------------------------------
   The stellar sample used for this work comes from the Hypatia catalog, a resource containing multiple literature sources of precise stellar abundances matched to \textit{Gaia} or \citet{Anderson_2012} stellar properties \citep{Hinkel_2014, Hinkel_2016}. From the catalog, we limit the sample to F, G, and K-type stars, with a \textit{V} mag < 11, and a log \textit{g} > 3.5 dex. We limit the stellar sample to simulate a potential PLATO sample that could be subject to high precision planetary mass-radius measurements \citep{Rauer_2014, Miglio_2017}. For this study, we selected stars with complete data on their absolute stellar abundances of Fe, Si, Mg, and O, as well as having an Mg/Si ratio as $0.7\leq$Mg/Si$\leq1.7$ due to model parametrization constraints (see Sect.~\ref{sec:modelb}). Throughout this paper, Fe/Si = $\mathrm{N}_{\mathrm{Fe}}/\mathrm{N}_{\mathrm{Si}}$ and Mg/Si =  $\mathrm{N}_{\mathrm{Mg}}/\mathrm{N}_{\mathrm{Si}}$, which are absolute stellar abundance elemental ratios.
   
   The catalog enables the user to differentiate between thin disc and thick disc stars, a distinction made by \citet{Hinkel_2014, Hinkel_2016} using the kinematics method from \citet{Bensby_2003}. We use the catalog's distinction to construct the samples of thin and thick disc stars. However, the thick disc sample appears to contain possible halo stars. To separate thick disc and halo stars, we employ the same procedure from \citet{Bensby_2003} and the criteria outlined in \citet{Bensby_2014}.
   
   The separation of the thick disc and halo stars is performed using a kinematically derived probabilistic measure of which galactic substructure a star belongs to. This separation method uses the stars’ galactic space velocity components (the radial velocity $U$, the rotational velocity $V$, and the vertical velocity $W$) compared to the characteristic substructures' galactic velocity dispersions as well as the observed populations' fraction of stars in the solar neighborhood \citep{Bensby_2003,Bensby_2014}.
   
\begin{figure}[!ht]
   \centering
   \includegraphics[width=\hsize]{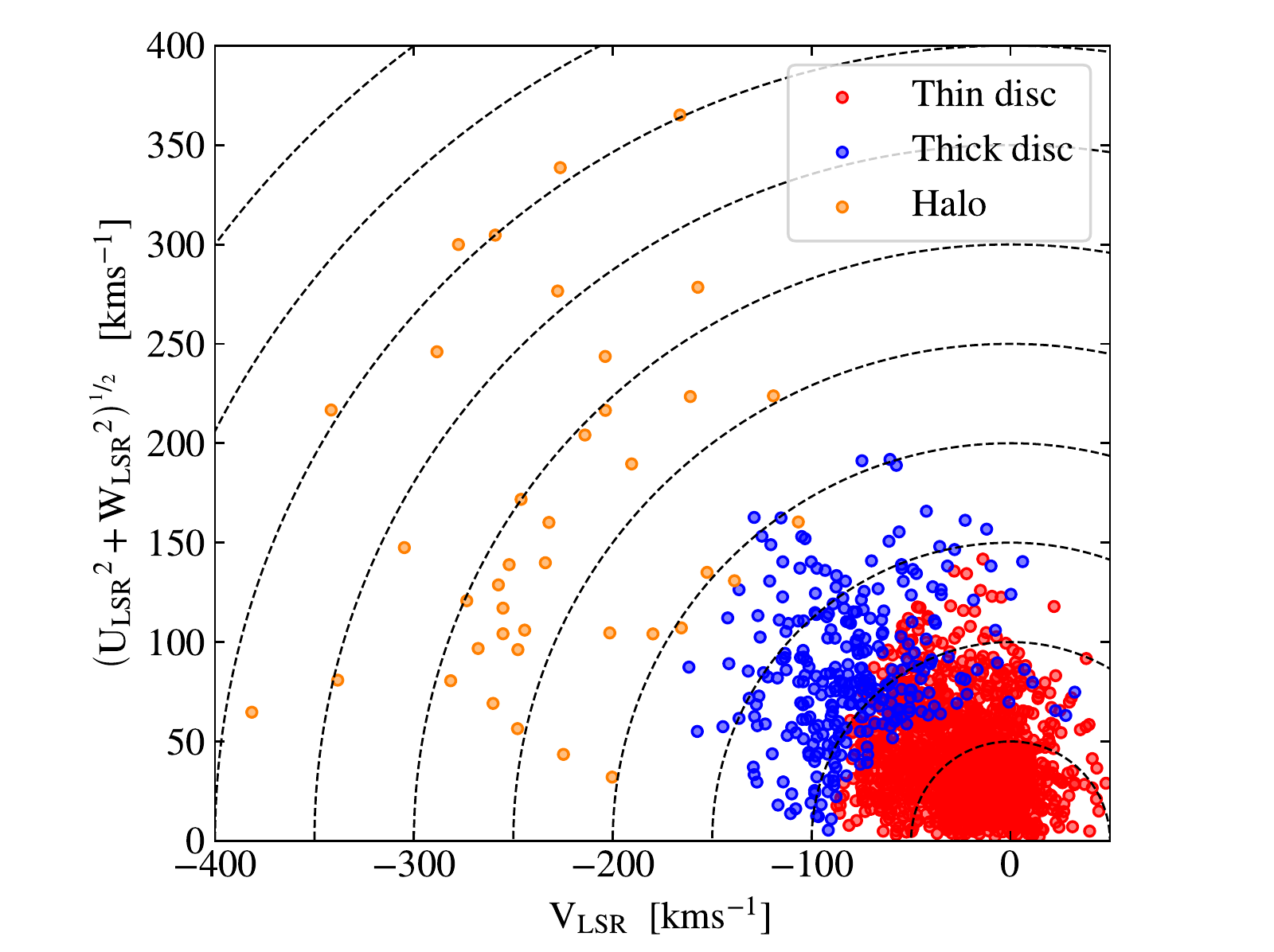}
      \caption{Toomre diagram demonstrating the stellar sample differentiated into the stellar populations. The red dots are likely thin disc stars, blue dots are likely thick disc stars, and orange dots are likely halo stars.} \label{fig:toomre}
   \end{figure}

   \citet{Bensby_2003} assume the galactic space velocities of the different stellar populations to have Gaussian distributions, 
   \begin{equation}
       f(U, V, W)=k \cdot \exp \left(-\frac{U^{2}}{2 \sigma_{U}^{2}}-\frac{\left(V-V_{\mathrm{asym}}\right)^{2}}{2 \sigma_{V}^{2}}-\frac{W^{2}}{2 \sigma_{W}^{2}}\right),
        \label{eq:f(U,V,W)}
   \end{equation}
   where 
   \begin{equation}
       k=\frac{1}{(2 \pi)^{3 / 2} \sigma_{U} \sigma_{V} \sigma_{W}}
   \end{equation}
   is the normalization constant \citep{Bensby_2003}. $\sigma_{U}$, $\sigma_{V}$, and $\sigma_{W}$ are the characteristic velocity dispersions, and $V$\textsubscript{asym} is the asymmetric drift, the values of these variables can be found in \citet{Bensby_2003}. We then use \citet{Bensby_2003} method to calculate the thick-disc-to-halo (TD/H) relative probability,
   \begin{equation}
       \mathrm{TD/H} = \frac{X_{\mathrm{TD}}}{X_{\mathrm{H}}} \cdot \frac{f_{\mathrm{TD}}}{f_\mathrm{H}},
        \label{eq:TD/H}
   \end{equation}
   where $X$ is the observed fraction of stars in the Solar neighbourhood from \citet{Bensby_2003} and $f$ is calculated from the star's velocities, see Eq. (\ref{eq:f(U,V,W)}). When the value of TD/H $<1$ then it is more likely a halo star \citep{Bensby_2014} and when TD/H $>1$ it is more likely a thick disc star. We determine the final overall stellar sample to be consisting of 1\,949 thin disc stars, 295 thick disc stars, and 41 halo stars. In Fig.~\ref{fig:toomre} we plot the overall stellar sample in a Toomre diagram illustrating the general differentiation of the stellar populations according to their total space velocities. However, the Toomre diagram highlights how the stellar populations total space velocities overlap due to the additional probabilistic separation criteria employed to categorize the stars.
% ----------------------------------------------------------------------------------------
\section{Methods}
\label{sec:methods}
% ----------------------------------------------------------------------------------------
   We used two theoretical models to numerically compute an expected solid planet structure. Using different models allows us to quantify model uncertainties to check for potential biases coming from the models.
% ----------------------------------------------------------------------------------------
\subsection{Model A}
\label{sec:modela}
% ----------------------------------------------------------------------------------------
   Model A is based on the comprehensive planet evolution code (\verb$completo21$) \citep{Mordasini_2012}. We compute an expected planetary \textit{M - R} relationship, methodology from \citet{Mordasini_2012}. The planetary structure is calculated by solving the 1D internal structure equations \citep{Bodenheimer_1986},
   \begin{eqnarray}
      \partial M/\partial R & = & 4 \pi R^2 \rho    \label{eq:intstr-mass} \\
      \partial P/\partial R & = & -G M \rho / R^2, \label{eq:intstr-pres}
   \end{eqnarray}
   using a modified polytropic equation of state (EoS) from \citet{Seager_2007}, as following,
   \begin{equation}
     \rho(P) = \rho_0 + cP^n.
   \end{equation}
   The main materials within the model treatment include pure iron, silicates (perovskite, enstatite) and ice, whose parameters $\rho_0$, $c$ and $n$ are from \citet{Seager_2007}. The EoS used provides the density as a function of pressure for a wide pressure range, however, it neglects the impact of temperature on density, and on the radius. Although the impact on the radius is small \citep{Valencia_2007, Grasset_2009}, a first-order temperature correction is used on the mean density for the current model version (see Appendix A from \citet{Linder_2019}).
% ----------------------------------------------------------------------------------------
\subsection{Model B}
\label{sec:modelb}
% ----------------------------------------------------------------------------------------
   Model B is based on the forward model for planetary structure inference by \citet{Dorn_2015}. Modifications to \citet{Dorn_2015} include different equations of state and the calculation of the adiabatic gradient in each layer directly from the equation of state (Haldemann et al. (in prep)).
   
   The radius of a planet is computed solving the static 1D internal structure Eqs.~\ref{eq:intstr-mass} and~\ref{eq:intstr-pres} given a planet mass $M_{total}$, a bulk composition ($\mathrm{N}_{\mathrm{Fe}}$/$\mathrm{N}_{\mathrm{Si}}$, $\mathrm{N}_{\mathrm{Mg}}$/$\mathrm{N}_{\mathrm{Si}}$, and water mass fraction $f_{water}$), surface temperature, and pressure. The planet is split into three layers, an iron core of mass $M_{core}$, a silicate mantle of mass $M_{rock}$, and a layer of water on the surface of mass $M_{water}$ (this latter condition is only included in the scenario where we consider a water layer). We use the EoS of \citet{Sotin_2007} for the mantle and the water layer; in the core for pressures above 240~GPa we use the EoS of \citet{Hakim_2018} and for lower pressures, we use the EoS of \citet{Fei_2016}. The temperature profile in each layer is assumed to be adiabatic as,
   \begin{equation}   
     \frac{\partial T}{\partial R} = \frac{\partial P}{\partial R} \frac{P}{T} \left(\frac{\mathrm{d} \ln T}{\mathrm{d} \ln P}\right)_S.
   \end{equation}
   The choice of silicate phases in the mantle description of \citet{Sotin_2007} restricts the possible mantle compositions to values which fulfill the relation
   \begin{equation}
     1 \leq \left(\mathrm{N}_{\mathrm{Fe}}/\mathrm{N}_{\mathrm{Si}}\right)_{rock} + \left(\mathrm{N}_{\mathrm{Mg}}/\mathrm{N}_{\mathrm{Si}}\right)_{rock} \leq 2,
   \end{equation}
   where $\left(\mathrm{N}_{\mathrm{Fe}}/\mathrm{N}_{\mathrm{Si}}\right)_{rock}$ and $\left(\mathrm{N}_{\mathrm{Mg}}/\mathrm{N}_{\mathrm{Si}}\right)_{rock}$ are the elemental ratios in the rocky mantle. Assuming $\left(\mathrm{N}_{\mathrm{Fe}}/\mathrm{N}_{\mathrm{Si}}\right)_{rock} = 0$ as in model A would cause a large fraction of cases to be outside of the parametrization laid out by \citet{Sotin_2007}. Therefore, we adopt a value of $\left(\mathrm{N}_{\mathrm{Fe}}/\mathrm{N}_{\mathrm{Si}}\right)_{rock} = 0.3$ to minimize the number of cases outside the parametrization of \citet{Sotin_2007}. The assumed value of 0.3 for the iron to silicate mantle ratio is in between the expected Earth and Mars ratios \citep{Dreibus_1987}.
% ----------------------------------------------------------------------------------------
\subsection{Model inputs}
\label{sec:inputs}
% ----------------------------------------------------------------------------------------
   We employed part of the stoichiometric model from \citet{Santos_2017} where the stellar abundances of Fe, Si, Mg, C, O, He, and H are used to derive the important rock types and chemical species for the composition of terrestrial-like planets \citep{Seager_2007, Sotin_2007}. We use the iron mass fraction and water mass fraction as inputs for model A, thus, the pertinent rock types and species from the \citet{Santos_2017} model are MgSiO\textsubscript{3}, Mg\textsubscript{2}SiO\textsubscript{4}, SiO\textsubscript{2}, Fe, and H\textsubscript{2}O. We disregard the species used by \citet{Santos_2017} to calculate the summed mass percent of all heavy elements. 
   
   We calculate, according to \citet{Santos_2017} stoichiometric model, the iron mass fraction $\left(f_{\mathrm{iron}}\right)$ of the rocky silicate-iron core and water mass fraction $\left(f_{\mathrm{water}}\right)$ present in the planetary bulk composition. We thus use the following relations from \citet{Santos_2017}, 
   \begin{eqnarray}
        f_{\text{iron}}=m_{\mathrm{Fe}} /\left(m_{\mathrm{Fe}}+m_{\mathrm{MgSiO} 3}+m_{\mathrm{Mg} 2 \mathrm{SiO} 4}+m_{\mathrm{SiO} 2}\right) \\
        f_{\text{water}}=m_{\mathrm{H} 2 \mathrm{O}} /\left(m_{\mathrm{H} 2 \mathrm{O}}+m_{\mathrm{Fe}}+m_{\mathrm{Mg} \mathrm{SiO} 3}+m_{\mathrm{Mg} 2 \mathrm{SiO} 4}+m_{\mathrm{SiO} 2}\right), \label{eq:fwater}
   \end{eqnarray}
   where $m_X=N_X\cdot \mu_X$, $N_X$ represents the number of atoms of each species $X$, and $\mu_X$ is their mean molecular weights; all $N_X$ values are computed relative to hydrogen \citep{Santos_2017}. We use part of the stoichiometric relations outlined in Appendix B of \citet{Santos_2017} to find the $N$ values of the different chemical species used in the mass fractions. As inputs to the models for mass fractions, we use \citet{Santos_2017} stoichiometric relations,
   \newline\noindent when $N_{\mathrm{Mg}} > N_{\mathrm{Si}}$,
   \begin{eqnarray}
        N_{\mathrm{O}}&=&N_{\mathrm{H}_{2} \mathrm{O}} + 3 N_{\mathrm{MgSiO}_{3}} + 4 N_{\mathrm{Mg}_{2} \mathrm{SiO}_{4}} \\ 
        N_{\mathrm{Mg}}&=&N_{\mathrm{MgSiO}_{3}} + 2 N_{\mathrm{Mg}_{2} \mathrm{SiO}_{4}} \\ N_{\mathrm{Si}}&=&N_{\mathrm{MgSiO}_{3}} + N_{\mathrm{Mg}_{2} \mathrm{SiO}_{4}}, 
    \end{eqnarray}
   alternatively, if $N_{\mathrm{Mg}} \leq N_{\mathrm{Si}}$,
   \begin{eqnarray} 
        N_{\mathrm{O}}&=&N_{\mathrm{H}_{2} \mathrm{O}} + 3 N_{\mathrm{MgSiO}_{3}} + 2 N_{\mathrm{SiO}_{2}} \\
        N_{\mathrm{Mg}}&=&N_{\mathrm{MgSiO}_{3}} \\ N_{\mathrm{Si}}&=&N_{\mathrm{MgSiO}_{3}} + N_{\mathrm{SiO}_{2}}.
    \end{eqnarray}
   From these, the iron mass fraction and water mass fraction are calculated for every star in the stellar sample.

\begin{table}[!ht]
      \caption{Average values and standard deviations of iron mass fraction ($f_{\mathrm{iron}}$) and water mass fraction ($f_{\mathrm{water}}$) percentages expected in solid planets of different galactic populations. The $f_{\mathrm{water}}$ is considered only for solid planets forming past the ice line, prior to such, we consider 0\% $f_{\mathrm{water}}$ for all three stellar populations.}  
      \label{table:iron_water_mass}
      \centering
      \begin{threeparttable}
         \begin{tabular}{l c c}
            \hline
            \noalign{\smallskip}
            &  $f_{\mathrm{iron}}$  (\%) & $f_{\mathrm{water}}$  (\%) \\
            \hline
            \noalign{\smallskip}
            This work &  & \\
            \noalign{\smallskip}
            Thin disc & $30.72 \pm3.76$ & $59.71 \pm9.52$\\
            Thick disc & $23.93 \pm4.68$ & $69.85 \pm8.63$\\
            Halo & $21.24 \pm4.03$ & $75.58 \pm6.20$\\
            \hline
            \noalign{\smallskip}
            \citet{Santos_2017} &  & \\
            \noalign{\smallskip}
            Thin disc & $31.974 \pm1.750$ & $59.713 \pm7.106$\\
            Thick disc & $24.305 \pm1.623$ & $72.179 \pm5.961$\\
            Halo & $23.110 \pm2.884$ & $83.990 \pm4.115$\\
            \hline
            \noalign{\smallskip}
            \citet{Cabral_2019}\tnote{1} &  & \\
            \noalign{\smallskip}
            Thin disc & $30.0 \pm0.8$ & $58.1 \pm0.4$\\
            Thick disc & $23.5 \pm1.7$ & $61.6 \pm0.9$\\
            Halo & $20.4 \pm0.2$ & $63.0 \pm0.1$\\
            \hline
            \noalign{\smallskip}
            Sun \citep{Lodders_2003} &  31.62 & 57.47 \\
            \hline
            \noalign{\smallskip}
         \end{tabular}
         \begin{tablenotes}
            \item[1] For comparison we select the Milky Way simulation from \citet{Cabral_2019}.
         \end{tablenotes}
      \end{threeparttable}
   \end{table}   
   
   We compare the resulting water and iron mass fractions from the stellar populations in our sample with \citet{Santos_2017} and \citet{Cabral_2019} (see Table~\ref{table:iron_water_mass}, distribution plots for our stellar sample can be found in Appendix~\ref{appendix:a}, Figs.~\ref{ironfrac} and~\ref{waterfrac}). We note that there is good agreement between the different overall stellar samples regardless as to the whether the stellar populations were distinguished by chemical \citep{Santos_2017, Cabral_2019} or kinematic constraints (this work). The iron mass and water mass fraction trends are conserved across each set of results. We report that the thin disc population has the highest iron mass fraction and the halo population has the lowest iron mass fraction; whilst the water mass fraction is the lowest in the thin disc and the highest in the halo stars. 
   
   For model B, we use absolute stellar abundance elemental ratios of Fe/Si and Mg/Si as well as the water mass fraction, the latter is calculated as outlined by the procedure above using \citet{Santos_2017} methodology. Within the elemental ratios, the Fe/Si variability is relatively distinct between the stellar populations; the highest ratio is in the thin disc and the lowest in the halo. This is expected and correlates with the iron mass fraction trends seen in Table~\ref{table:iron_water_mass}. A more detailed discussion pertaining to input distributions and distinctiveness between stellar populations is found in Appendix~\ref{appendix:a}. The Mg/Si ratio is not significantly variable between stellar populations and does not allow us to distinguish between stellar populations (see Figs.~\ref{fesi} and~\ref{mgsi}).
   
   Both models necessitate surface temperature inputs; for our purposes we calculate a planetary equilibrium temperature as a function of distance from the host star (Eq.~\ref{temp}), repeating this procedure for every star. We use the following relation to calculate the planetary equilibrium temperature, $T_{eq}$ as,
   \begin{equation}
       T_{eq}=\left(\frac{L_{\star}\left(1-A_{b}\right)}{16 \sigma \pi a^{2}}\right)^{1 / 4},
       \label{temp}
   \end{equation}
   where $L_\star$ is the stellar luminosity, $A_{b}$ is the planet albedo which we fix at 0.3, $\sigma$ is the Stefan-Boltzmann constant, and $a$ is the planet's orbital distance from its host star. We select planetary orbital distances of 0.1 and 4 AU. 
   For each star, an iron-mass fraction, water-mass fraction, Fe/Si, Mg/Si, and an expected planetary equilibrium temperature are calculated as inputs to the models. The variability in stellar abundances within one stellar population is expected to provide a distribution of radii per fixed mass in the range of 1 to 10 $M_{\oplus}$. We run both models at each fixed mass for the whole samples of each stellar population; this allows for a median of the planetary radii distribution to be calculated per stellar population per fixed mass to find the typical \textit{M - R} relationship. 

% ----------------------------------------------------------------------------------------
\subsection{Validation}
\label{sec:validation}
% ----------------------------------------------------------------------------------------
   We tested these models against a composition predicted by the solar photosphere chemical abundances from \citet{Lodders_2003} and a 0.01\% water mass fraction to account for current estimates of water on Earth. Running the models for a 1 $M_{\oplus}$ planet at an orbital distance of 1 AU from the Sun, model A yields a planet of 0.961 $R_{\oplus}$ and model B yields a planet of 1.013 $R_{\oplus}$. The two models differ systematically in the radius given the solar derived Earth composition. As such, we do not delve into a comparison of individual radii values between the models but, rather, how the different stellar populations' chemical composition impact the expected relative planetary radii differences between the stellar populations.
% ----------------------------------------------------------------------------------------
\section{Results}
\label{sec:results}
% ----------------------------------------------------------------------------------------
   \begin{figure*}
      \includegraphics[width=\textwidth]{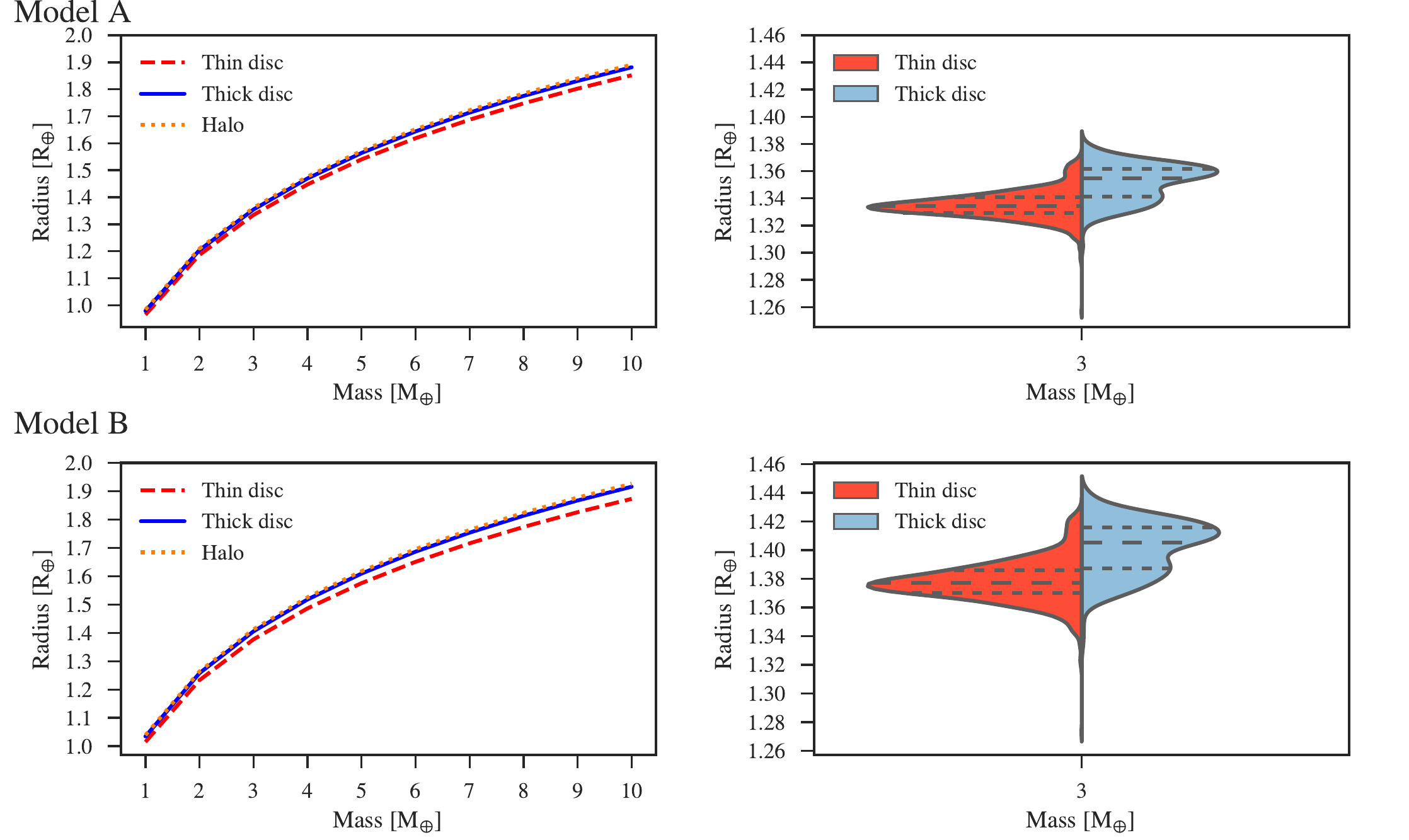}
      \cprotect\caption{At 0.1 AU, the expected \textit{M - R} relationships of silicate-iron planets in different galactic stellar populations are in the left panels and violin plots demonstrating expected radii distributions at 3 $M_{\oplus}$ are in the right panels. The two top panels are results obtained using model A and the two bottom panels are from model B.}
         \label{fig:01au}
   \end{figure*}
   \begin{figure*}        
      \includegraphics[width=\textwidth]{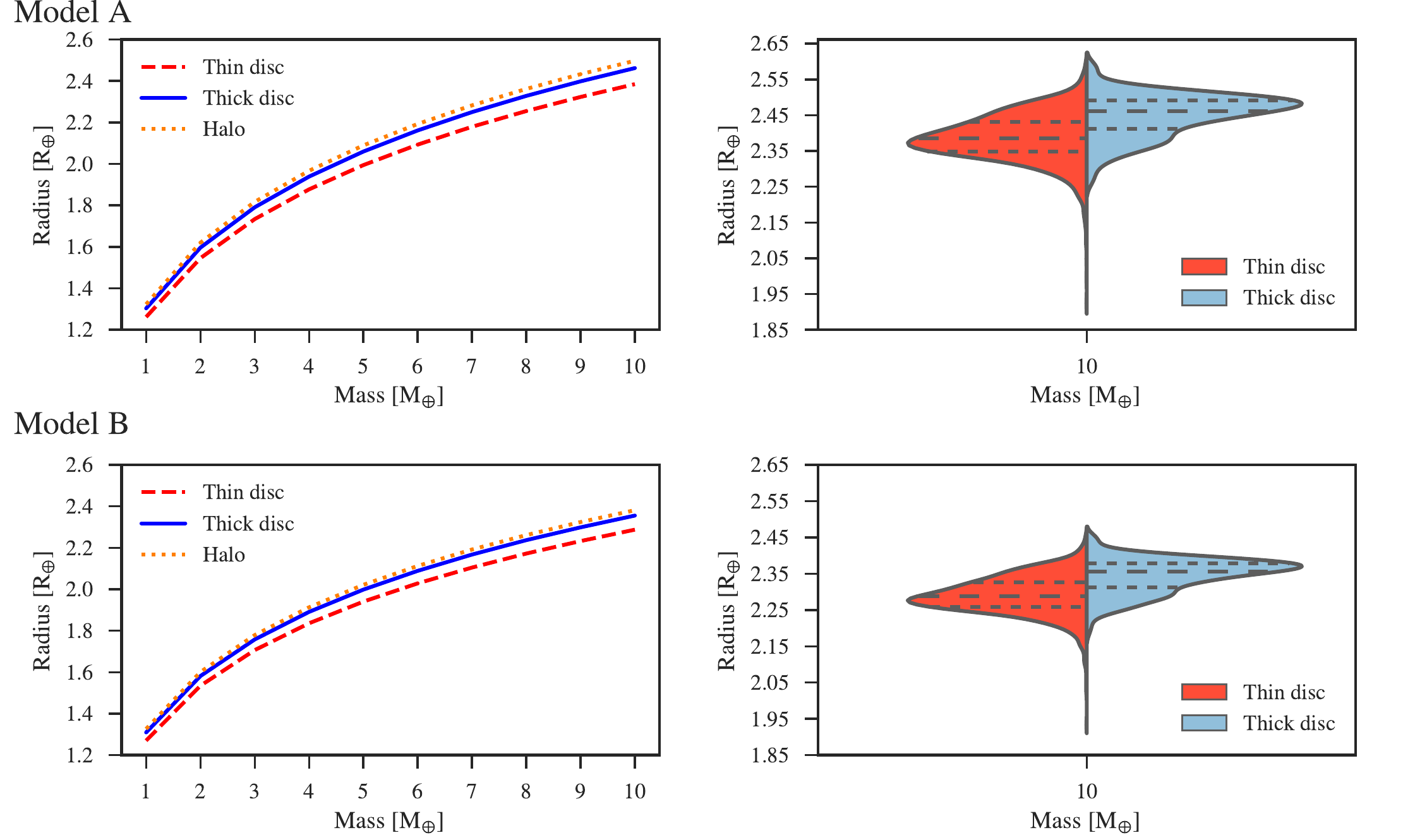}
      \cprotect\caption{Same plots as in Fig.~\ref{fig:01au} calculated at 4 AU for silicate-iron-water planets. The water mass fraction is now included to form a water ice layer around the rocky silicate-iron core. The violin plots in the right panels illustrate the expected radii distributions for 10 $M_{\oplus}$ planets.}
         \label{fig:4au}
   \end{figure*}
% ----------------------------------------------------------------------------------------   
   We present the planetary \textit{M - R} relationships for synthetic solid planets for a mass range of 1 to 10 $M_{\oplus}$ computed at 0.1 AU for a silicate-iron composition (Fig.~\ref{fig:01au}) and at 4 AU for a silicate-iron-water composition (Fig.~\ref{fig:4au}) by both models A and B. In the left panels of Figs.~\ref{fig:01au} and~\ref{fig:4au}, each stellar population yields different median solid planet \textit{M - R} relationships; the dashed red lines are the expected typical solid planets orbiting around thin disc stars; the solid blue lines are planets around thick disc stars; the dotted orange lines are planets around halo stars. In the right panels of Figs.~\ref{fig:01au} and~\ref{fig:4au}, violin plots at 3 $M_{\oplus}$ and 10 $M_{\oplus}$, respectively, illustrate the radii distributions of the solid planets for the thin and thick disc stellar populations. At 3 $M_{\oplus}$ at 0.1 AU and 10 $M_{\oplus}$ at 4 AU, for the three stellar population inputs, both models produce expected median planetary radii that are currently abundant in solid exoplanet observations \citep{Fulton_2017}.
   
   The two cases, silicate-iron planets at 0.1 AU and silicate-iron-water planets at 4 AU from their host star, were chosen to demonstrate two types of solid planets. The first case investigates a rocky silicate-iron core with no atmosphere and the ulterior case investigates a rocky silicate-iron core with a significant water ice layer. At 0.1 AU the expected planetary equilibrium temperature is on average between 850 to 880~K depending on the stellar population. Due to these high temperatures, no volatiles are taken into considerations for the scenario at 0.1 AU since they would have likely been evaporated. Instead, we focus on the effect of the stellar populations' iron content, placed in the rocky silicate-iron core, on the \textit{M - R} relationships. Seen in Fig.~\ref{fig:01au}, for both models, the thin disc stars with the highest iron-mass fraction and Fe/Si ratio yield the smallest typical planetary radii. The halo stars with the lowest iron-mass fractions and Fe/Si ratios yield planets with the largest typical radii at fixed masses on the \textit{M - R} diagram. 
   
   Since we use the whole sample of stars per stellar populations as inputs to the models, we have a distribution of planetary radii per fixed mass and per stellar population. There is variability in the stellar photospheric abundances within each sample stellar population (see Figs.~\ref{ironfrac},~\ref{waterfrac},~\ref{fesi}, and~\ref{mgsi}), thus there is a distribution of expected planetary radii as illustrated by the violin plots in the right panels of Figs.~\ref{fig:01au} and~\ref{fig:4au}. 
   
   At 0.1 AU the expected planetary radii distributions for a 3 $M_{\oplus}$ solid planet are displayed in the violin plots in Fig.~\ref{fig:01au}. This mass was chosen since it produces median radii between 1.334 to 1.413 $R_{\oplus}$ thereby being part of the abundant observations of transiting  super-Earths \citep{Fulton_2017}. Noticeably, the planetary radii distributions between the thin and thick disc stellar populations overlap. However, the median radius, as indicated by the thick dashed lines, demonstrate the difference in the expected typical radius of the different stellar populations. At 3 $M_{\oplus}$, we report a 1.48\% difference between the thick disc (1.354 $R_{\oplus}$) and thin disc (1.334 $R_{\oplus}$) median planetary radii and a 0.52\% between the halo (1.361 $R_{\oplus}$) and thick disc planetary radii using model A. As for model B, we calculate a 2.04\% radius difference between the thick disc (1.404 $R_{\oplus}$) and thin disc (1.377 $R_{\oplus}$), and a 0.66\% difference between the halo (1.413 $R_{\oplus}$) and thick disc. For planets at 0.1 AU, we find that there is an expected 1.51\% mean planetary radius difference between the thick and thin disc across the 1 to 10 $M_{\oplus}$ planets from model A and a 2.04\% radius difference from model B. Between the halo and thick disc, the mean planetary radius difference for 1 to 10 $M_{\oplus}$ is 0.53\% from model A and 0.69\% from model B. 

   In Fig.~\ref{fig:4au} we present the \textit{M - R} relationships for solid planets with a water ice layer, namely silicate-iron-water planets. The median planetary equilibrium temperatures computed for the stellar samples vary between 135 and 140~K, thereby condensing the refractory species as well as volatiles. We concern ourselves only with H\textsubscript{2}O \citep{Dorn_2017_2}. We also consider that all the H\textsubscript{2}O present in the protoplanetary disc is condensed during the accretion process. In the inputs, the water mass fraction is largest in the halo stars and lowest in the thin disc stars; this results in compounding the iron mass fraction variability to form denser planets around thin disc stars compared to thick disc and halo stars. From the \textit{M - R} relationships in Fig.~\ref{fig:4au}, the smallest planets orbit around thin disc stars and the largest planets around halo stars at fixed mass. At 10 $M_{\oplus}$ the models compute planets of radii comparable to \citet{Fulton_2017} abundant sub-Neptunes and \citet{Grasset_2009} 10 $M_{\oplus}$ solid planets with 60 to 80\% water mass fractions. In the violin plots of Fig.~\ref{fig:4au} we show the distributions of planetary radii. The median radius difference between the thick disc (2.461 $R_{\oplus}$) and thin disc (2.384 $R_{\oplus}$) is of 3.24\%, whilst between the halo (2.497 $R_{\oplus}$) and thick disc, it is a difference of 1.46\% using model A. With model B, we get similar results, there is a 2.88\% difference in median planetary radii between the thick disc (2.353 $R_{\oplus}$) and the thin disc (2.297 $R_{\oplus}$), and a 1.20\% difference between the halo (2.381 $R_{\oplus}$) and the thick disc. Overall for the selected 1 to 10 $M_{\oplus}$ range at 4 AU, we find that there is an expected 3.26\% radius difference between the thick and thin disc from model A and a 2.93\% radius difference from model B. Between the halo and thick disc, the mean planetary radius difference for 1 to 10 $M_{\oplus}$ is 1.49\% from model A and 1.24\% from model B. 

   For model A and to a lesser extent in model B, at both orbital distances, there are secondary peaks in the thick disc planetary radii distributions within the violin plots per fixed mass. These double peaks in the thick disc radii distributions are to the secondary peaks present in the iron fraction and Fe/Si distributions. The input for model A, the thick disc iron mass fraction distribution, has a primary peak centered at 22.1\% and a secondary peak is identified at 29.9\% (see Fig.~\ref{ironfrac}). For model B, the thick disc Fe/Si input distribution also has a similar feature, the primary peak is at 0.57 and is accompanied by a secondary peak at 0.86 (see Fig.~\ref{fesi}). The secondary peaks in the thick disc input parameters aforementioned are caused by an overabundance of thick disc stars with high Fe/Si ratios, this is then propagated into the thick disc solid planet radii distributions as seen in the violin plots of Figs~\ref{fig:01au} and~\ref{fig:4au}. In Appendix~\ref{appendix:a}, a more detailed treatment is provided as to why these two thick disc parameter inputs demonstrate bimodal Gaussian distributions.
   
   From these results, we see a consistent statistical difference between the solid planets that would form around the different stellar populations; at a fixed mass the typical planets around thin disc stars have smaller radii than planets around thick disc stars and halo stars for both cases at 0.1 and 4 AU (for a complete median radius and radius difference per stellar population and model at 0.1 AU, see Table~\ref{table:rad01AU}, and at 4 AU, see Table~\ref{table:rad4AU}). 
% ----------------------------------------------------------------------------------------
   \begin{figure}[!h]
      \includegraphics[width=\hsize]{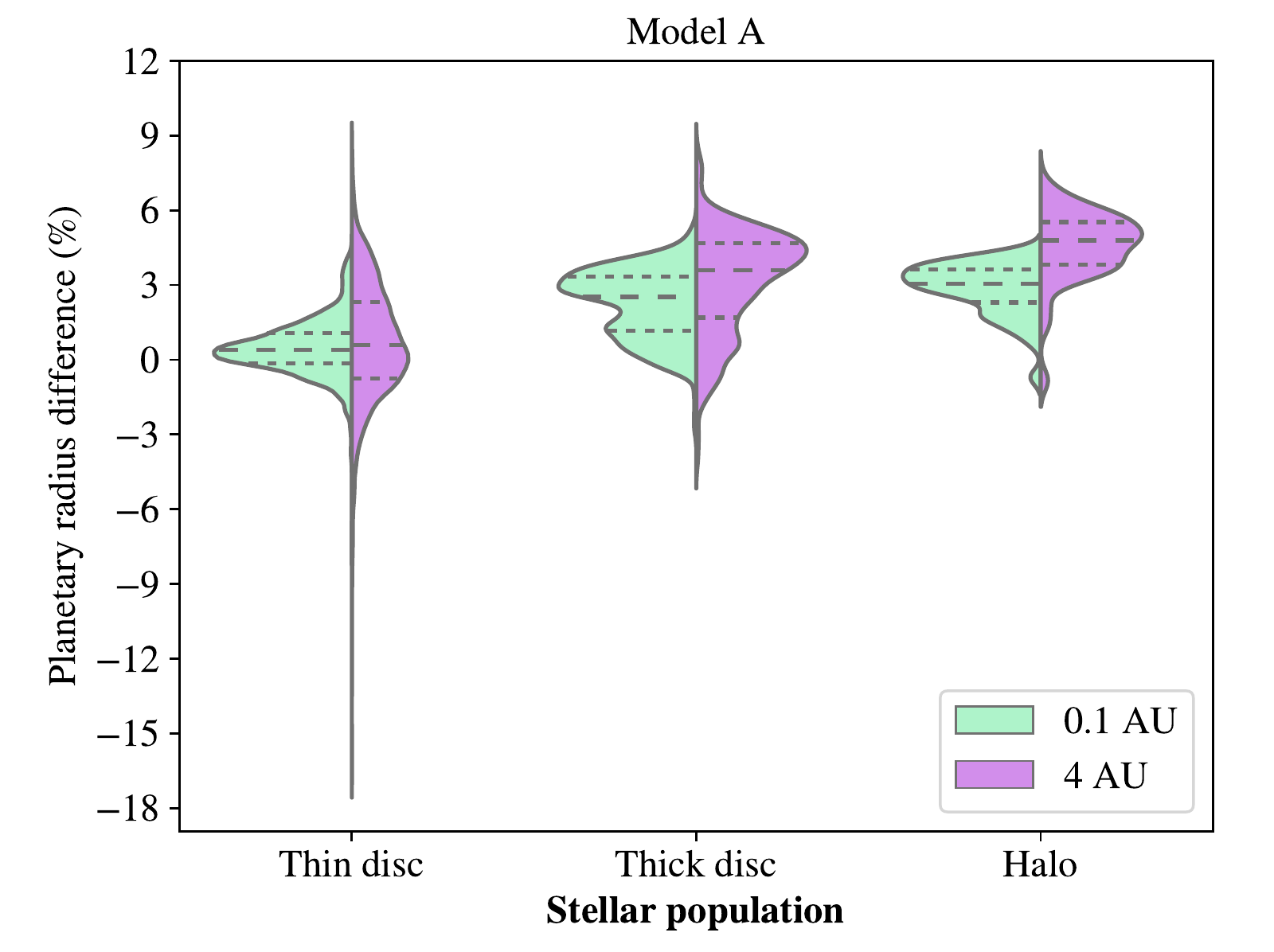}
      \cprotect\caption{Mass aggregate comparison of the planetary radii distribution percentage difference from the solar-derived expected planets at 0.1 and 4 AU as computed by model A.}\label{fig:modela}
   \end{figure}
   \begin{figure}[!h]       
      \includegraphics[width=\hsize]{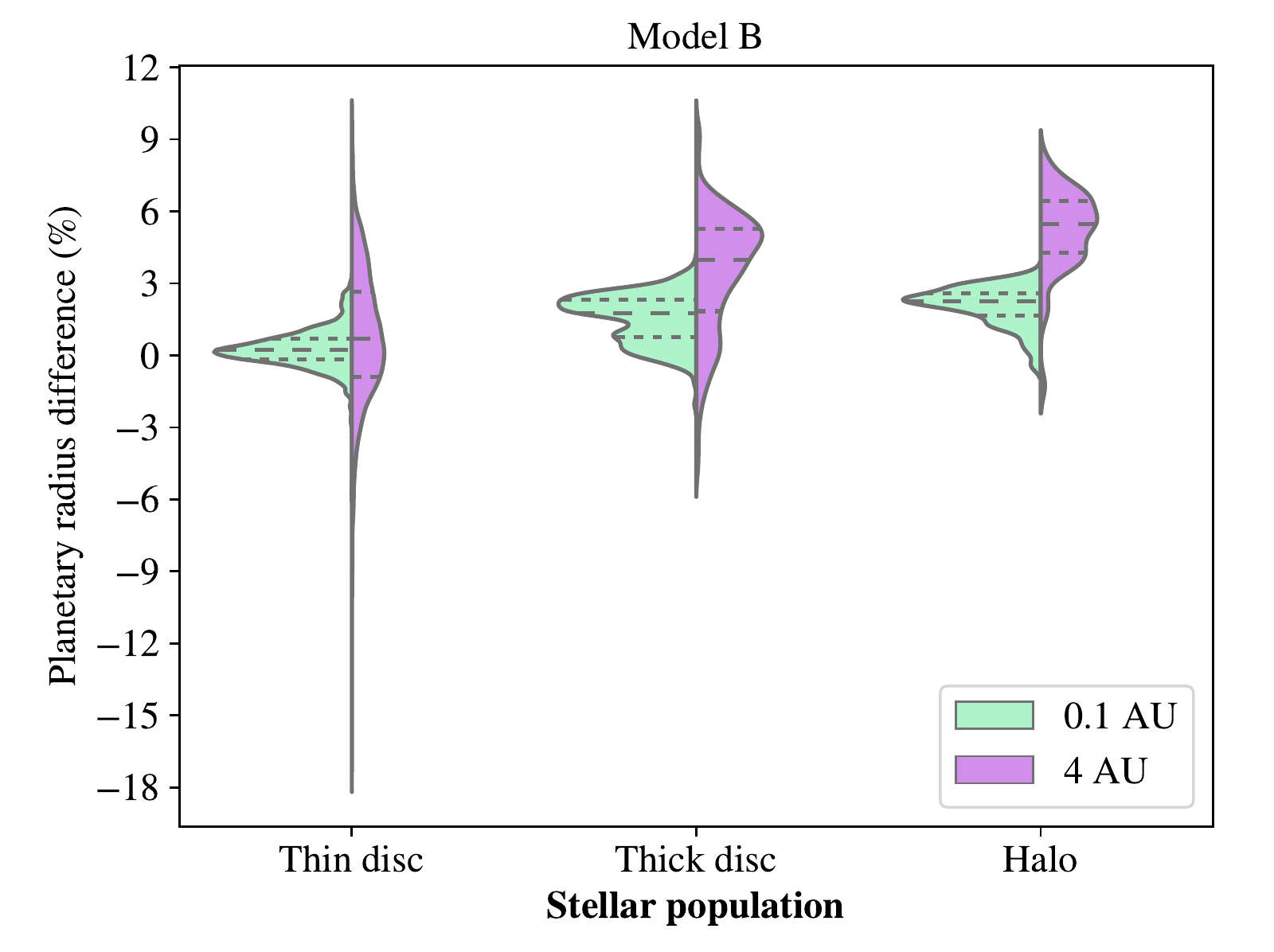}
      \cprotect\caption{Same plots as in Fig.~\ref{fig:modela} using model B.}\label{fig:modelb}
   \end{figure}
% ----------------------------------------------------------------------------------------  

   To compare the three stellar population radii distributions, we benchmark these against the expected model-calculated solar system planets at 0.1 and 4 AU using the methodology outlined in Sec.~\ref{sec:inputs} applied to \citet{Lodders_2003} solar photospheric abundances. The range of planet masses within the stellar populations are aggregated and compared to the expected solar planets at 0.1 and 4 AU from 1 to 10 $M_{\oplus}$. The violin plots in Figs.~\ref{fig:modela} and~\ref{fig:modelb} illustrate the planetary radius difference per stellar population compared to the solar-derived planet benchmark for models A and B, respectfully. From these, the planetary radii percentage difference distributions are seen to differ between the stellar populations across the aggregate mass range. The planets orbiting around halo stars have a larger percentage difference from the solar benchmark than those around thin disc or thick disc stars. Across stellar populations and models, the presence of the water ice layer in the silicate-iron-water planets at 4 AU increases the planetary radii compared to the dry silicate-iron planets at 0.1 AU.

% ----------------------------------------------------------------------------------------
\section{Discussion and Outlook}
\label{sec:discussion}
% ----------------------------------------------------------------------------------------
   The results obtained from the kinematically separated galactic stellar populations demonstrate that there is a physical difference in the solid planet \textit{M - R} relationships between the thin disc, thick disc, and halo populations. For both models at 0.1 AU, the relative radii differences between the stellar populations increase with mass. However, at 4 AU, for both models, an increase in mass results in a decrease in the relative radii differences between the stellar populations. This is due to the larger compressibility of the water ice layer at higher pressures and thus reducing the difference in radii at larger planetary masses with larger water masses.
   
   The stellar abundance variability between the galactic stellar populations was demonstrated by \citet{Santos_2017}, \citet{Cabral_2019}, and confirmed again with our different stellar sample and separation method. We show that the chemical distinctiveness of the various galactic stellar populations can be partially retrieved after probabilistic kinematic differentiation for our initial stellar sample from the Hypatia catalog, see Appendix~\ref{appendix:a}. 
   
   The bulk composition of solid exoplanets impacts the \textit{M - R} relations as demonstrated by \citet{Mordasini_2012} and \citet{Dorn_2017_2} when testing internal structure models. Thereby from the stellar-derived bulk planetary compositions, we confirm these results as being coherent with the literature \citep{Mordasini_2012, Dorn_2017_2, Santos_2017, Cabral_2019}. The planetary \textit{M - R} relationship differences between the stellar populations are demonstrated using two different models and are persistent for both dry silicate-iron planets at 0.1 AU and water-rich silicate-iron-water planets at 4 AU. The two models, as seen in Sec.~\ref{sec:validation}, for a solar-derived Earth composition, do not compute an exact 1 $R_ {\oplus}$ but rather show a systematic difference. As such, the exact planetary radii calculated for the three stellar populations is not the core focus of this work, rather we aim to show the systematic median planetary radii differences between the stellar populations with two separate models.
   
   This work rests on the basic assumption that the stellar photosphere's chemical composition approximates solid planets' bulk compositions well. Many studies have provided evidence that this approximation is reasonable when examining the refractory species' abundances in chondritic meteorites, Earth, Venus, and Mars compared to our Sun's photosphere \citep{Morgan_1980, McDonough_1995, Drake_2002, Lodders_2003, Khan_2008}. This presumption, that stellar abundance is a good proxy for solid planet composition, is currently widely used in planetary bulk composition calculations and planet internal structure models \citep{Santos_2015, Thiabaud_2015, Brugger_2017, Santos_2017, Cabral_2019, Wang_2019}. The iron mass fraction relationship between the host star and planet was further examined by \citet{Santos_2015} where they show that stars hosting exoplanets, CoRoT-7, Kepler-10, and Kepler-93, have photospheric abundances that are thought to be reflected in their transiting rocky planets. In contrast, Mercury and dense planets, i.e. K2-106 b or K2-229 b, who are often very close in to their host stars, have demonstrated larger iron cores amounting up to possibly 80\% of the planetary mass, thereby differing considerably from their host star's composition \citep{Guenther_2017, Santerne_2018}.
   
   The present-day solid planet composition may vary from the expected initial bulk composition due to processes happening in the protoplanetary disc during formation and after disc dispersal. Deviations in the planetary bulk composition could be the result of fractionation and incomplete condensation \citep{Wasson_1974}, partial volatile evaporation \citep{Braukmuller_2018, Lichtenberg_2019}, varying accretion and differentiation mechanisms \citep{Fischer_2017}, initial location, potential migration in the protoplanetary disc and migration post disc \citep{Mordasini_2012, CarterBond_2012}, and collisional stripping post disc formation \citep{Marcus_2010}. These possible evolutionary complications during planet formation are hindrances to using the stellar photospheric abundances as proxies for solid planet compositions. These impacts are believed to be less significant for the refractory species inclusion, as a scenario at 0.1 AU, but are tantamount to the water mass fraction as in our scenario at 4 AU.
   
   At 4 AU, in both models, we use a pure water ice layer as a substitute for all other ice types (commonly these include CO, CO\textsubscript{2}, CH\textsubscript{4}, NH\textsubscript{3}). According to \citet{Dorn_2017_2}, this substitution is reasonable since O is more abundant than C and N, and water condenses at higher temperatures than other ice species. Another minor limitation comes from our decision to not compute any atmosphere neither at 0.1 nor at 4 AU. Furthermore, we do not take into account the stellar abundance uncertainties, their impact is negligible on the final solid planet radii.
   
   In regards to the planetary equilibrium temperatures calculated at 0.1 and 4 AU, we do not use bolometric corrections between the different star types and rather just make a general statement of the expected temperature at these orbital distances. The temperature is a rough estimate used to control the type of chemical species that condense into the bulk planetary composition at the selected orbital distances. We use this to determine whether to include the water mass fraction as calculated by Eq.~(\ref{eq:fwater}) or to assume negligible amounts of water are present and use a water mass fraction of 0.
   
   The results show a solid planet \textit{M - R} difference across the two computational models for a range of fixed masses from 1 to 10 $M_{\oplus}$ at two separate orbital distances of 0.1 and 4 AU from their host star. The key idea can be seen in the statistics of the distributions, it cannot be used as a stand alone result with individual exoplanet characterizations within these stellar populations. 
   
   To confirm these results, observations of a statistically important number of solid exoplanets are required, particularly for thick disc and halo stars. Presently, observers have mainly characterized planets around thin disc stars. According to \citet{Adibekyan_2012_2}, for given sub-solar metallicities, the planet occurrence rate is thought to be greater in the thick disc compared to the thin disc. Additionally, \citet{Bashi_2019} provide evidence that the small-planet occurrence rate is high for both iron-poor high velocity stars such as thick disc ones, and iron-rich low velocity stars, thin disc stars. There have already been observations of some thick disc stars hosting exoplanets, namely Kepler-10 and Kepler-444 \citep{Dumusque_2014, Campante_2015} as well as a halo star, Kapteyn \citep{Anglada-Escude_2014} and more observations can be expected in the future. As new space-based telescopes (including PLATO) precisely characterize solid planets in various stellar populations, planetary \textit{M - R} relations will be better constrained. From precise planetary \textit{M - R} measurements we will also be able to confirm or disbar the idea of using the stellar photosphere composition as a proxy for solid planets' bulk compositions. Lastly, our result may help to assess the potential habitability of planets around stars in different stellar populations.
% ----------------------------------------------------------------------------------------
\begin{acknowledgements}
      A.M. acknowledges the opportunity to conduct research at the Physikalisches Institut, Universität Bern. J.H. acknowledges the support from the Swiss National Science Foundation under grant 200020$\_$172746. C.M. acknowledges the support from the Swiss National Science Foundation under grant BSSGI0$\_$155816 ``PlanetsInTime''. Parts of this work have been carried out within the frame of the National Center for Competence in Research PlanetS supported by the SNSF. We thank the referee for their constructive comments and suggestions.
\end{acknowledgements}
% ----------------------------------------------------------------------------------------
\bibliographystyle{aa}
\bibliography{main}
% ----------------------------------------------------------------------------------------
\begin{appendix}
% ----------------------------------------------------------------------------------------
\section{Stellar populations input composition distributions}
\label{appendix:a}
\begin{figure}[!h]
   \centering
   \includegraphics[width=\hsize]{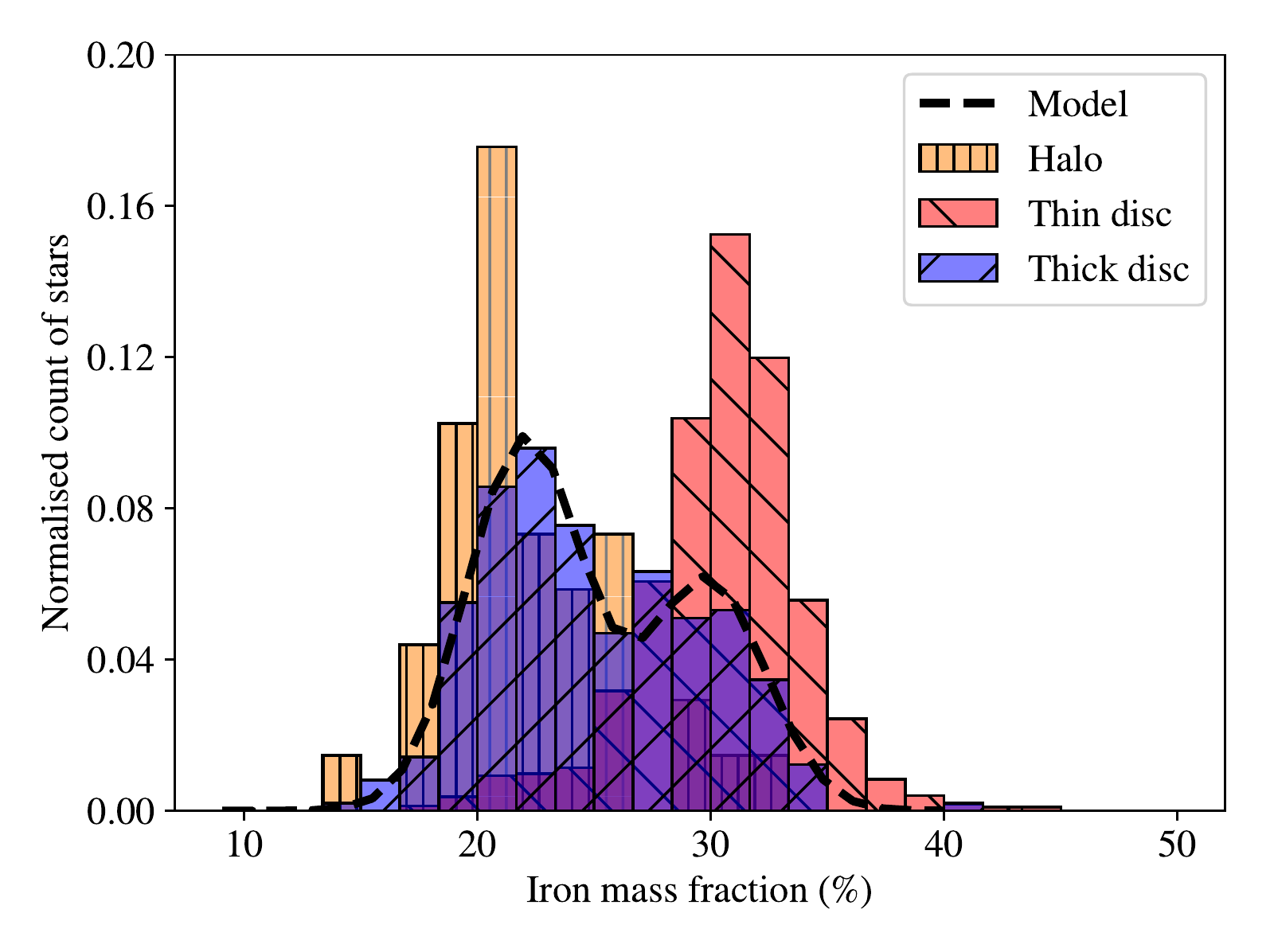}
      \caption{Normalized histogram illustrating the iron mass fraction distributions of the three kinematically separated stellar populations. The thick disc iron mass fraction distribution is bimodal and fitted by two Gaussians with a primary peak, $f_{\mathrm{iron}}\sim22.1\%$ and a standard deviation $\sigma_{f_{\mathrm{iron}}}\sim2.5$ and a secondary peak, $f_{\mathrm{iron}}\sim29.9\%$ with $\sigma_{f_{\mathrm{iron}}}\sim2.5$.}
         \label{ironfrac}
   \end{figure}

\begin{figure}[!h]
   \centering
   \includegraphics[width=\hsize]{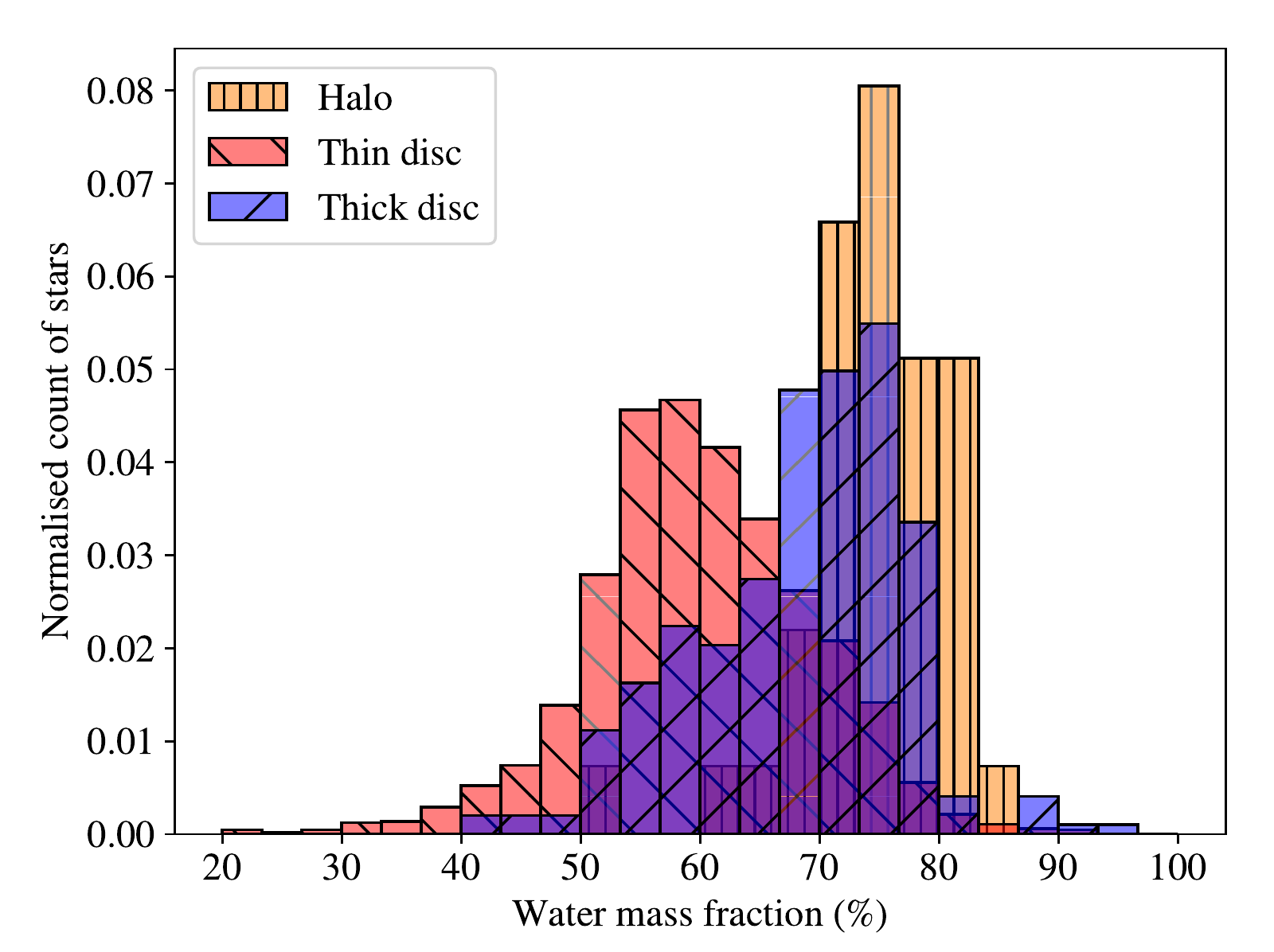}
      \caption{Normalized histogram illustrating the water mass fraction distributions of the three kinematically separated stellar populations.}
         \label{waterfrac}
   \end{figure}

\begin{figure}[!h]
   \centering
   \includegraphics[width=\hsize]{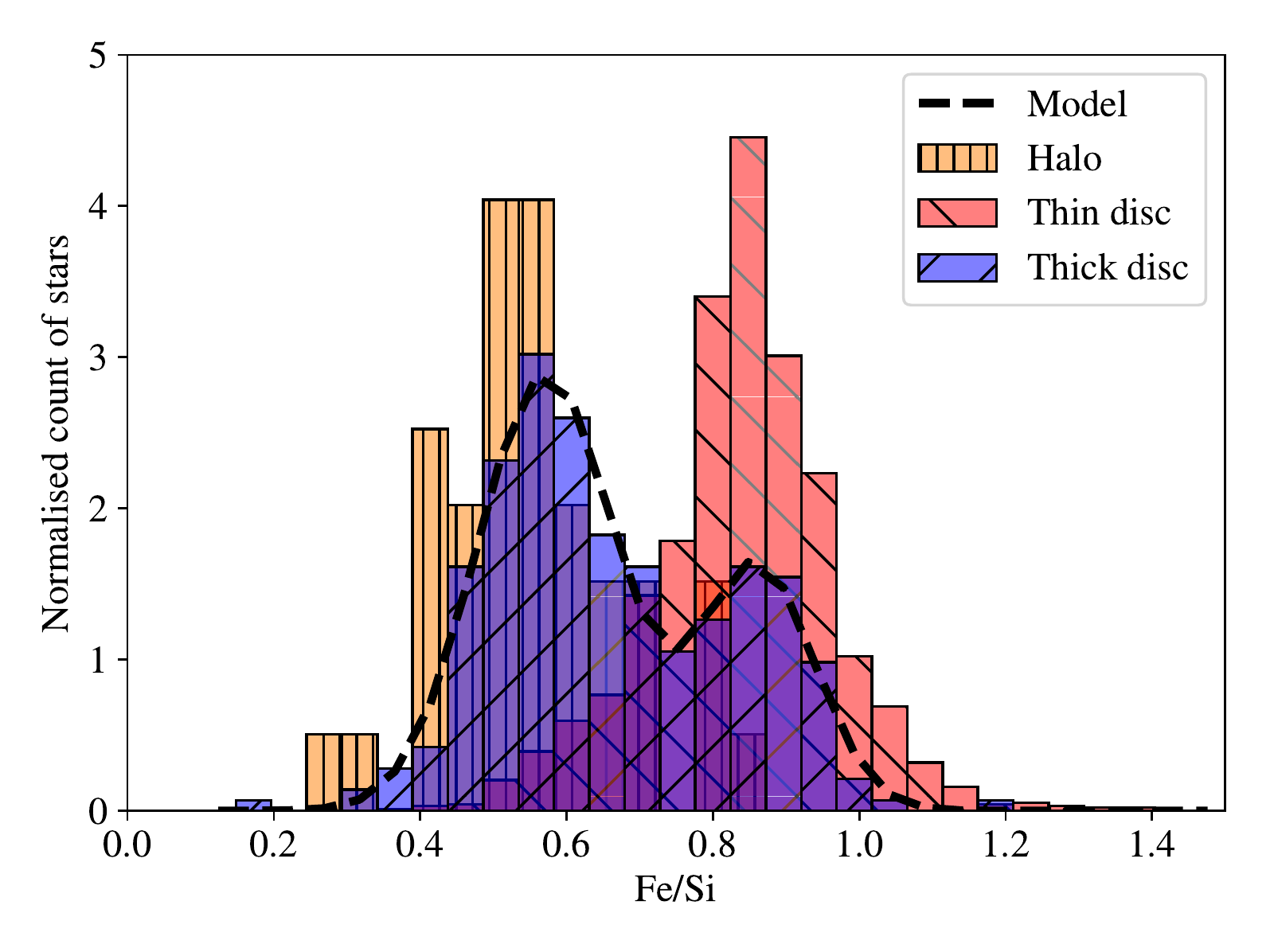}
      \caption{Normalized histogram illustrating the Fe/Si ratio distributions of the three kinematically separated stellar populations. The thick disc Fe/Si distribution is bimodal and fitted by two Gaussian with a primary peak, $\mathrm{Fe/Si}\sim0.57$ and $\sigma_{\mathrm{Fe/Si}}\sim0.09$ and a secondary peak, $\mathrm{Fe/Si}\sim0.86$ with $\sigma_{\mathrm{Fe/Si}}\sim0.08$.}
         \label{fesi}
   \end{figure}

\begin{figure}[!h]
   \centering
   \includegraphics[width=\hsize]{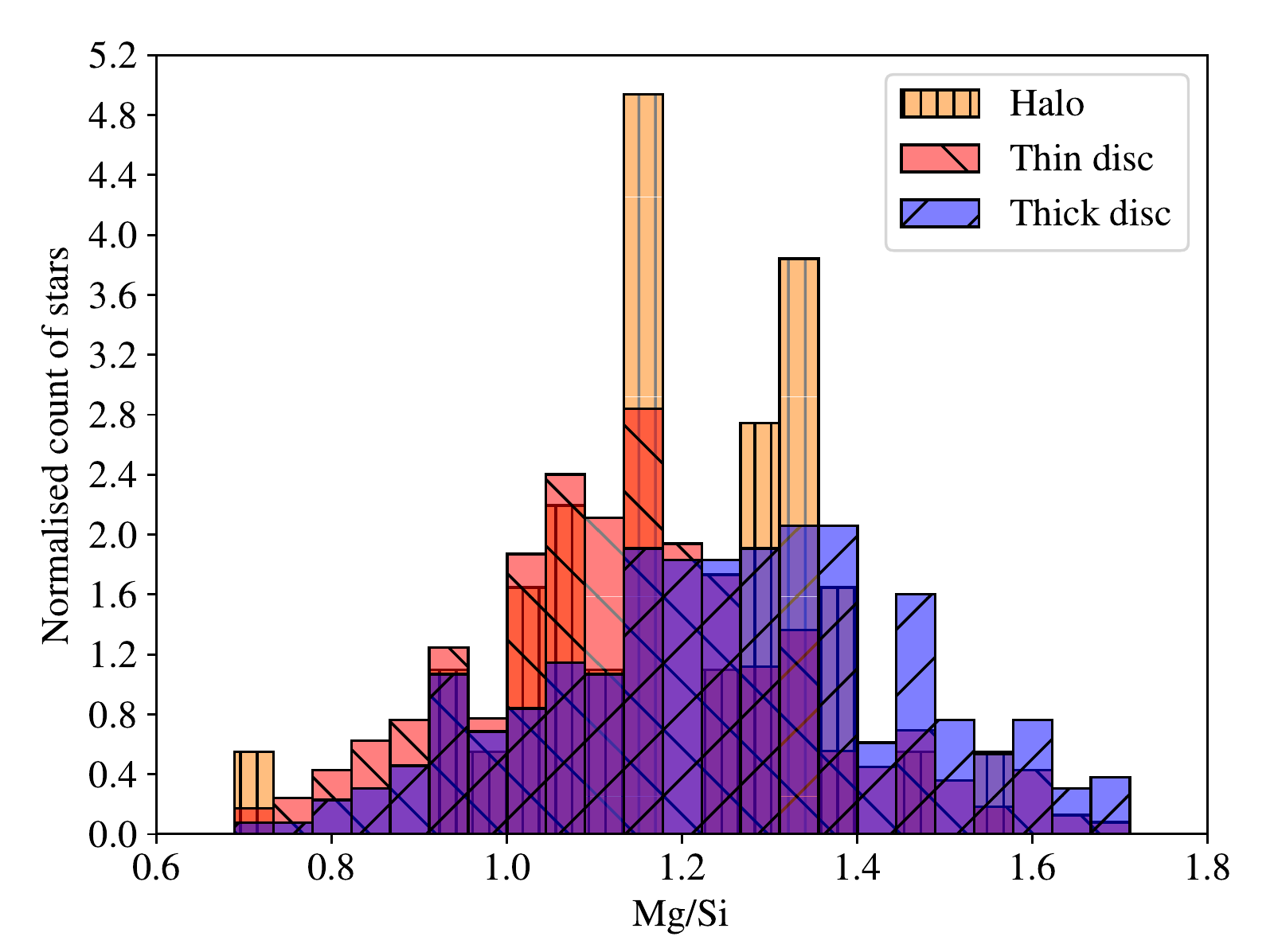}
      \caption{Normalized histogram illustrating the Mg/Si ratio distributions of the three kinematically separated stellar populations.}
         \label{mgsi}
   \end{figure}

% ----------------------------------------------------------------------------------------
   The distribution inputs to the two models are displayed in Figs.~\ref{ironfrac},~\ref{waterfrac},~\ref{fesi}, and~\ref{mgsi}, where we differentiate the stellar populations according to the probabilistic kinematic method presented in Sec.~\ref{sec:data}. The input compositions to the two models, iron mass fraction, water mass fraction, Fe/Si, although being differentiable in their average composition between stellar populations as presented in Table~\ref{table:iron_water_mass}; they are not completely distinct and overlap; in particular, the Mg/Si distributions between populations is not differentiable.
   
   We highlight the presence of a secondary peak in the iron mass fraction and Fe/Si thick disc distributions as seen and plotted in Figs.~\ref{ironfrac} and~\ref{fesi}, respectively. We explain such a secondary feature in the thick disc distributions to be caused by an overabundance of high Fe/Si thick disc stars. This comes from the potential biases in the original stellar sample, which is not a volume limited one, but rather an assortment of catalogs of stars within 150 pc. This collection of stars is further constrained according to our data completeness requirements as outlined in Sec.~\ref{sec:data}. We thus have a stellar sample of thick disc stars with an excess of high Fe/Si ratios which we do not expect to see within other large magnitude limited surveys such as LAMOST \citep{Yan_2019}.
   
% ----------------------------------------------------------------------------------------
\section{Radii and differences per mass per stellar region}
\label{appendix:b}
% ----------------------------------------------------------------------------------------
   Below we provide the median \textit{M - R} values for masses from 1 to 10 $M_{\oplus}$, for both models and at both orbital distances, at 0.1 AU dry silicate-iron planets in Table.~\ref{table:rad01AU} and at 4 AU the silicate-iron-water planets in Table.~\ref{table:rad4AU}. In addition, per mass and model, we calculate the expected solid planet median radii difference in percent between the thick disc $\left(R_{\mathrm{Thick}}\right)$ and thin disc $\left(R_{\mathrm{Thin}}\right)$, as being,
   \begin{equation*}
    \mathrm{Thick-Thin} = \frac{R_{\mathrm{Thick}}-R_{\mathrm{Thin}}}{R_{\mathrm{Thin}}} \cdot 100
   \end{equation*}
   and between the halo $\left(R_{\mathrm{Halo}}\right)$ and thick disc as,
   \begin{equation*}
    \mathrm{Halo-Thick} = \frac{R_{\mathrm{Halo}}-R_{\mathrm{Thick}}}{R_{\mathrm{Halo}}} \cdot 100.
   \end{equation*}
   These serve to demonstrate how the radius differences between the stellar populations evolve as we increase in mass but also how these compare between the two models. In the last row of both tables we calculate the mean percentage radius difference across the mass range between the stellar populations.
   {\renewcommand{\arraystretch}{1.3}
    \begin{table*}
    \caption{Detailed median radii per mass per model for planets that form at 0.1 AU from their host star. For the models the inputs for the water mass fraction is 0. The median radius per stellar population is shown in earth radii $R_{\oplus}$ and the median radius difference between different stellar populations is calculated as a percentage.}             
    \label{table:rad01AU}      
    \centering          
    \begin{tabular}{| c | c c c | c c | c c c | c c |}
    \hline
     & \multicolumn{5}{|c|}{Model A} & \multicolumn{5}{|c|}{Model B}\\ \cline{2-11}
    & \multicolumn{3}{|c|}{Median Radius ($R_{\oplus}$)} & \multicolumn{2}{|c|}{Radius difference (\%)} & \multicolumn{3}{|c|}{Median Radius ($R_{\oplus}$)} & \multicolumn{2}{|c|}{Radius difference (\%)} \\ \cline{2-11}
    Mass ($M_{\oplus}$) & Thin & Thick & Halo & Thick-Thin & Halo-Thick & Thin & Thick & Halo & Thick-Thin & Halo-Thick \\
    \hline
    1 & 0.965 & 0.979 & 0.984 & 1.43 & 0.50 & 1.015 & 1.033 & 1.039 & 1.80 & 0.62 \\  
    2 & 1.186 & 1.204 & 1.210 & 1.46 & 0.51 & 1.233 & 1.256 & 1.264 & 1.88 & 0.66 \\
    3 & 1.334 & 1.354 & 1.361 & 1.48 & 0.52 & 1.377 & 1.404 & 1.413 & 1.96 & 0.66 \\
    4 & 1.447 & 1.469 & 1.477 & 1.50 & 0.53 & 1.486 & 1.516 & 1.526 & 2.00 & 0.68 \\
    5 & 1.540 & 1.563 & 1.572 & 1.51 & 0.53 & 1.575 & 1.607 & 1.618 & 2.03 & 0.69 \\
    6 & 1.619 & 1.643 & 1.652 & 1.52 & 0.54 & 1.651 & 1.684 & 1.696 & 2.06 & 0.70 \\  
    7 & 1.687 & 1.713 & 1.722 & 1.53 & 0.55 & 1.716 & 1.752 & 1.764 & 2.11 & 0.70 \\
    8 & 1.748 & 1.775 & 1.784 & 1.54 & 0.55 & 1.774 & 1.811 & 1.824 & 2.23 & 0.73 \\
    9 & 1.802 & 1.830 & 1.840 & 1.54 & 0.55 & 1.825 & 1.865 & 1.879 & 2.15 & 0.74 \\
    10 & 1.852 & 1.881 & 1.891 & 1.55 & 0.56 & 1.873 & 1.913 & 1.927 & 2.16 & 0.75 \\
    \hline
    Mean & & & & 1.51 & 0.53 & & & & 2.04 & 0.69 \\
    \hline                  
    \end{tabular}
    \end{table*}   
    }
    {\renewcommand{\arraystretch}{1.3}
     \begin{table*}
     \caption{Detailed median radii per mass per model for planets that form at 4 AU from their host stars. The water mass fraction is extracted from the stoichiometric model outlined in Sec.~\ref{sec:inputs}. The median radius per stellar population is shown in earth radii $R_{\oplus}$ and the median radius difference between different stellar populations is calculated as a percentage.}  
     \label{table:rad4AU}      
     \centering          
     \begin{tabular}{| c | c c c | c c | c c c | c c |}
     \hline
     & \multicolumn{5}{|c|}{Model A} & \multicolumn{5}{|c|}{Model B}\\ \cline{2-11}
     & \multicolumn{3}{|c|}{Median Radius ($R_{\oplus}$)} & \multicolumn{2}{|c|}{Radius difference (\%)} & \multicolumn{3}{|c|}{Median Radius ($R_{\oplus}$)} & \multicolumn{2}{|c|}{Radius difference (\%)} \\ \cline{2-11}
     Mass ($M_{\oplus}$) & Thin & Thick & Halo & Thick-Thin & Halo-Thick & Thin & Thick & Halo & Thick-Thin & Halo-Thick \\
     \hline
    1 & 1.261 & 1.303 & 1.323 & 3.31 & 1.53 & 1.270 & 1.309 & 1.326 & 3.05 & 1.35 \\ 
    2 & 1.545 & 1.596 & 1.620 & 3.28 & 1.51 & 1.535 & 1.580 & 1.601 & 2.98 & 1.31 \\
    3 & 1.733 & 1.790 & 1.817 & 3.26 & 1.50 & 1.706 & 1.757 & 1.779 & 2.94 & 1.27 \\
    4 & 1.877 & 1.938 & 1.967 & 3.26 & 1.49 & 1.836 & 1.890 & 1.913 & 2.94 & 1.24 \\
    5 & 1.994 & 2.059 & 2.089 & 3.25 & 1.48 & 1.940 & 1.997 & 2.021 & 2.92 & 1.24 \\
    6 & 2.092 & 2.160 & 2.192 & 3.25 & 1.48 & 2.028 & 2.087 & 2.112 & 2.90 & 1.22 \\ 
    7 & 2.178 & 2.249 & 2.282 & 3.25 & 1.47 & 2.104 & 2.165 & 2.192 & 2.89 & 1.21 \\
    8 & 2.254 & 2.327 & 2.361 & 3.24 & 1.47 & 2.172 & 2.234 & 2.261 & 2.87 & 1.20 \\
    9 & 2.322 & 2.398 & 2.433 & 3.24 & 1.47 & 2.232 & 2.297 & 2.324 & 2.89 & 1.20 \\
    10 & 2.384 & 2.461 & 2.497 & 3.24 & 1.46 & 2.297 & 2.353 & 2.381 & 2.88 & 1.20 \\ 
    \hline
    Mean & & & & 3.26 & 1.49 & & & & 2.93 & 1.24 \\
     \hline                  
     \end{tabular}
     \end{table*}   
    }
\end{appendix}
% ----------------------------------------------------------------------------------------
\end{document}